\newcommand{\InGaN}{{In$_x$Ga$_{1-x}$N}}
\documentclass[showkeys,showpacs,prb,aps,twocolumn,superbib]{revtex4}

\usepackage{graphicx}
\usepackage{color}


\begin{document}
\title{
First-principles calculation of the thermodynamics of
In$_x$Ga$_{1-x}$N alloys:  \\
Lattice vibrations effects
}
\author{Chee Kwan Gan$^1$, Yuan Ping Feng$^2$, and David~J.~Srolovitz$^3$}
\affiliation{
$^1$Institute of High Performance Computing, 1 Science Park Road, 
Singapore 117528, Singapore\\
$^2$Department of Physics, National University of Singapore, 2 Science Drive 3,
Singapore 117542, Singapore \\
$^3$Department of Mechanical and Aerospace Engineering, 
Princeton University, Princeton, New Jersey, 08544
} 

\begin{abstract}

The thermodynamics properties of the wurtzite and zinc-blende \InGaN\
alloys are calculated using first-principles density-functional
calculations. Special quasi-random structures are used to describe the
disordered alloys, for $x= 1/4, 1/2$, and $3/4$.
The effect of lattice vibrations on the phase diagram, commonly omitted
from semiconductor alloy phase diagram calculations, are included
through first-principles calculations of phonon spectra. 
Inclusion of lattice vibrations leads to a large reduction in the 
order-disorder critical temperature ($\sim 29$\% and $\sim 26$\% for
the wurtzite and zinc-blende structures, respectively) and changes the shape
of the solubility and spinodal curve through changes in the entropies 
of the competing phases. Neglect of such effect produces significant errors
in the phase diagrams of complex ordered semiconductor compounds.
The critical 
temperature for phase separation is 1654~K (1771~K) 
for the wurtzite (zinc-blende) structures.  The predicted phase diagrams are
in agreement with experimental measurements on MOCVD \InGaN\
films. 
\smallskip
\end{abstract}
\maketitle

\section{Introduction}
The group-III nitrides GaN and InN and their alloys have been
receiving considerable attention for high-power, high-frequency, and
high-temperature optoelectronic applications, such as light emitting
and laser diodes.  By changing composition one can, in
principle, continuously tune the band gap from 0.8~eV (InN)\cite{Nanishi03v42} 
to 3.5~eV
(GaN) --- a range that spans much of the visible spectrum.  However,
because of the large lattice mismatch between GaN and InN ($\sim11\%$
for the $a$ and $c$-directions in the wurtzite
structure), the solid solution has a tendency to undergo phase separation.
\cite{Osamura75v46,El-Masry98v72,Doppalapudi98v84}
GaN, InN and their alloys (\InGaN), grown under ambient
conditions, typically assume the hexagonal wurtzite (WZ) crystal
structure.\cite{Singh97v70,Doppalapudi98v84,El-Masry98v72,Sato97v36}
It is possible, however, to grow thin epitaxial films of \InGaN\ with
the cubic zinc-blende (ZB)
structure.\cite{Paisley89v7,Brandt95v52,Schikora96v54,Yang99v74,Gamez-Cuatzin99v176,Taniyasu00v180,Okumura98v189,Tabata99v75}
The phase stability of \InGaN\ alloys has been the subject of several
experimental\cite{El-Masry98v72,Doppalapudi98v84} and theoretical studies\cite{Ho96v69,Saito99v60,Takayama00v88,
vanSchilfgaarde97v178,Adhikari04v95,Karpov04v70,Matsuoka97v71,
Teles00v62,Ferhat02v65,Purton05v15}. 
For example, there have been more than ten calculations of the
\InGaN\ pseudo-binary phase diagram.
These have included valence-force field
(VFF)\cite{Ho96v69,Saito99v60,Takayama00v88}, tight-binding, and
density functional calculations\cite{vanSchilfgaarde97v178}, sometimes
combined with atomic-scale Monte Carlo
simulations\cite{Adhikari04v95}.  Surprisingly, there is considerable
discrepancy in the predicted miscibility gap.  The tendency for phase
separation is largely driven by the relatively large misfit
strains\cite{Teles00v62}. The predicted critical temperature
varies over a 2000~K
range, with most calculations predicting $1500{\rm~K} \le T_c \le
2000{\rm~K}$.  The source of this discrepancy may be associated with
the accuracy of the description of the atomic interactions, the need
for a statistically meaningful description of the distribution of Ga
and In on the solid solution cation sites and the contribution to the
free energy associated with configurational entropy\cite{Karpov04v70}.
Additionally, the vibrational contribution to the free energy has
never been included.  In the present paper, we report the results of
the most rigorous calculation of this phase diagram performed to date.
The results clearly demonstrate the importance of the vibrational
contribution in determining phase diagrams in semiconductor alloys, such as \InGaN.

In this work, we employ accurate density-functional theory (DFT) to examine
the thermodynamic properties of both wurtzite and zinc-blende \InGaN.  We directly
compare the formation enthalpy and free energy differences determined
using exactly the same pseudopotentials and energy cutoff for both
structures.  Using these results, we predict the binodal and spinodal
on the binary phase diagram; delineating the miscibility gap and the
region where the solid solution is unstable.  The simulation method
uses the special quasirandom structure (SQS)
formalism\cite{Zunger90v65,Wei90v42} to faithfully (and efficiently)
represent the structure of the random alloy.  The effect of lattice
vibrations, often neglected in semiconductor alloy thermodynamic
calculations, is included in a harmonic approximation where the
dynamical matrix is determined using density-functional theory
methods. Even though lattice
vibration effects have been shown to be significant for the
thermodynamic properties of some intermetallics\cite{vandeWalle02v74},
the effects of vibration have been commonly neglected for semiconductor alloys.
Such effects were not previously included, in part,
because an accurate treatment of lattice vibrations requires
substantial computing resources, above that needed for modeling the
effects of alloying on the heat of formation.  Advances in
first-principles computational methods and hardware make
routine inclusion of vibrational effects in alloy phase diagram
calculations possible.  As we demonstrate below, lattice vibrations
can drastically alter calculated thermodynamic properties and phase
diagrams of \InGaN.
This work provides the first direct evaluation of the significance of
lattice vibration effects on the phase diagram of compound
semiconductor alloys --- a large and technologically important class
of complex alloys. Earlier work on solid solution semiconductor alloys
Si-Ge\cite{Garbulsky96} and Ga$_{1-x}$In$_{x}$P\cite{Silverman95v7} suggested that lattice vibration effects are
negligible.  On the other hand, inclusion of these effects on metallic
compounds (intermetallics) have been shown to be important in
determining phase stability in the Cd-Mg\cite{Asta93v48},
Al-Sc\cite{Ozolins01v86}, Cu-Au\cite{Ozolins98v58}, and
Al-Cu\cite{Wolverton01v86} systems. However, vibration effects have been shown
to be relatively unimportant in other metallic systems that exhibit compound formation, e.g.,
Ni-Al\cite{vandeWalle98v80} and Pd-V\cite{vandeWalle00v61}.  The
previous work on the phase stability of metals suggests that
vibrational effects may or may not be important in those systems.
Further, there is no evidence to support the idea that one can
extrapolate from metals to compound semiconductor alloys.  

The solid state  \InGaN\ phase diagrams, determined
herein, represent the most rigorous determinations reported to-date.
The resultant phase diagram shows that the wurtzite crystal structure
is more stable than the zinc-blende structure over the entire \InGaN\
composition and temperature ranges examined.

\section{Previous theoretical calculations}
Due to its low computation cost and relative accuracy,
valence-force-field (VFF) models\cite{Keating66v145} have been used
extensively to 
study\cite{Ho96v69,Saito99v60,Takayama00v88,Adhikari04v95} the \InGaN\
alloy system.  One of the first \InGaN\ phase diagram calculations was
performed by Ho and Stringfellow\cite{Ho96v69}, who employed a  
VFF model to investigate the solid phase miscibility gap
in zinc-blende \InGaN.  They found a critical temperature of 1523~K
and that at the growth temperature of 800$^{\circ}$C, the InN
solubility in GaN is less than 6\%.  Using interaction constants
deduced from first-principles calculations, Saito and
Arakawa\cite{Saito99v60} simulated the wurtzite \InGaN\ alloy system
using the VFF model and obtained a critical temperature of 1690~K.
Takayama {\it et al.}\cite{Takayama00v88}, using a modified VFF for \InGaN,
found a critical temperature of 1668~K for the
zinc-blende structure and 1967~K for the wurtzite structure.

Several first-principles calculations have also been performed.  Van
Schilfgaarde {\it et al.}\cite{vanSchilfgaarde97v178} relaxed a
32-atom structure using a tight-binding method and determined the heat
of formation using density-functional theory. Combining this
result with an ideal mixing estimate of the configurational entropy
led to critical temperatures of 3980~K and 2290~K for the wurtzite and
zinc-blende structures, respectively. These values are considerably
higher than those obtained from the VFF models.
Teles {\it et al.}\cite{Teles00v62} investigated strained and relaxed
\InGaN\ alloys using a pseudopotential plane-wave approach with a
generalized quasichemical approximation (GQCA) for
mixing\cite{Sher87v36}. They obtained a critical temperature of 1295~K
for the zinc-blende \InGaN, which is lower than the VFF
results\cite{Ho96v69,Takayama00v88}.  Another first-principles
calculation\cite{Ferhat02v65} also reported a similar value for the
critical temperature (i.e., 1400~K) for zinc-blende \InGaN.

Grosse and Neugebauer\cite{Grosse01v63} assessed the limits and
accuracy of the VFF models for \InGaN\ alloys by studying various
ordered structures using first-principles calculations. They found
that while VFF works reasonably well for some zinc-blende III/V
semiconductors, significant errors occur for group III nitrides (GaN,
InN).  In particular, the formation energies for alloys are
significantly underestimated (by 14~meV/cation).  These
findings suggest that the critical temperature of \InGaN\
obtained from first-principles calculations should be higher than those
from VFF calculations. However, this conclusion is inconsistent with
the predictions of VFF\cite{Ho96v69,Saito99v60,Takayama00v88} and
first-principles results\cite{Teles00v62,Ferhat02v65} (i.e., the $T_c$
values from VFF studies exceed those from first-principles calculations). Our
results, presented below, shed light on this discrepancy.

\section{Special quasirandom structures}
\label{sec:sqs}
The effects of disorder on the stability and structure of alloys have
received considerable attention over the past few decades.  The
coherent-potential approximation (CPA) \cite{Ehrenreich76} for
describing such effects is a single-site theory that treats the random
A$_x$B$_{1-x}$ alloys by considering the average occupations of
lattice sites by atoms A and B. Hence the effects of local environment
such as charge transfer, local chemical environment, and local
structural relaxation are ignored. However, local environment effects
have been shown to be important in determining the thermodynamics and
electronic properties of alloys\cite{Zunger90v65,Wei90v42}.  One
approach to incorporate the local environment effect is through
first-principles density-functional theory.  One can, in principle,
model a disordered A$_x$B$_{1-x}$ system using a plane-wave
density-functional methods with a large supercell in which A and B are
distributed at random on lattice sites.  Unfortunately, since
computational costs in such methods usually scale with the number of
atoms in the supercell, $N$ as $N^3$, one is usually forced to use a
small supercell.  Small supercells invariably lead to poor statistics
and spurious correlations (i.e., this approach leads to a chemical
environment about each site which is different from that in a large,
truly random system).  Zunger {\it et al.}\cite{Zunger90v65,Wei90v42}
recognized that most physical properties are governed by atomic
interactions between near neighbors and developed a method to generate
atomic arrangements in small supercells that correctly reproduce the
short and intermediate range correlations that exist in random alloys.
These structures are called special-quasirandom-structures (SQS).

A useful approach to defining correlation functions in crystals with
different site occupancies is to assign each site $i$ a variable
according to the type of occupying atom, say $s_i=+1$ for an A atom
and $s_i=-1$ for an B atom.  Next, we make a list of pairs of atoms in
our structure that are separated by $k$ or less nearest neighbor
distances and evaluate the product $\pi=s_i s_j$ for that particular
pair of atoms on sites $i$ and $j$.  Averaging this value over all
such pairs in the structure yields the correlation function
$\Pi_{2,k}$, where the two indicates we are considering pairs.  We can
use a similar procedure for triplets, quadruplets,.. $m$-lets of atoms
to produce a series of correlation functions $\Pi_{m,k}$.  If the
alloy were of infinite extent and the site occupancies were truly
random, all $\Pi_{m,k}$ would be $\Pi_{m,k}(x)=\overline{\Pi}_m(x)=[(+1)x +
(-1)(1-x)]^m = (2x-1)^m$, where $x$ is the concentration of A atoms in
the system.

We have implemented the SQS algorithm for different unit cell sizes 
and shapes, generating all possible site occupancies for the chosen $x$,
calculating $\Pi_{m,k}$ and keeping the configuration for which the set of
$\Pi_{m,k}(x)$ is closest to $\overline{\Pi}_m(x)$.  In practice,
we use only $m=2$ and $k\le10$.  We repeat this procedure for
different unit cell shapes, as described by the supercell lattice
vectors ${\bf a}_1$, ${\bf a}_2$, and ${\bf a}_3$, and retain the one
that leads to the best correlation with random
occupancy (using our implementation of the procedure described in Refs. \onlinecite{Jiang04v69,vandeWalle03v26}).

In this work we have focused on wurtzite and zinc-blende crystal structures with 16
cations and 16 anions per unit cell.  The occupancies, in our case
correspond to Ga or In on the cation sites.  The descriptions of the
hexagonal-close-packed (hcp) and face-centered-cubic (fcc) sublattices
are shown in Tables~\ref{tab:wurtzitesqs16} and \ref{tab:zinc-blendesqs16},
respectively.  For $x=1/4$ (or $x=3/4$), the wurtzite structure delivers
exact matches up to $\Pi_{2,5}$ and up to $\Pi_{2,7}$ for $x=1/2$.
For the zinc-blende structure, we find structures that yield correlation
functions that exactly match the random case up to $\Pi_{2,3}$ for
$x=1/4$ (or $x=3/4$) and to $\Pi_{2,7}$ for $x = 1/2$.  Unlike in previous
studies of \InGaN\ alloys\cite{Ferhat02v65}, 
the present analysis provides a quantitative measure
of the degree to which the simulation cell resembles a random alloy.   
Based on the quoted pair statistics, we conclude that the unit cells and occupancies 
employed herein provide good representations of the set of wurtzite and zinc-blende
random alloys. 

\section{Formation Enthalpy Calculations}
\label{sec:dft}
We first optimize the zinc-blende (wurtzite) GaN and InN structures with respect to the
atomic coordinates and the unit cell parameter(s) $a$ (and $c$) using the
first-principles density-functional\cite{Hohenberg64v136,Kohn65v140}
method within the local density approximation (LDA).  This was done
using the plane-wave pseudopotential Vienna {\it ab initio} simulation
package (VASP)\cite{Kresse96v54}
non-norm-conserving pseudopotentials.\cite{Vanderbilt90v41} The
important\cite{Wright95v51,Grosse01v63} 3d states for Ga and 4d states
for In are included as valence electrons in the pseudopotentials.  A
relatively high cutoff energy of 31.96~Ry is used throughout this
work.  A Monkhost-Pack mesh of $7 \times 7 \times 7$ is used for the
4-atom unit cell wurtzite InN and  GaN systems. A 
$5\times 5 \times 5$ mesh is used for the 8-atom zinc-blende
InN and GaN unit cells  (finer meshes resulted in total energy changes of less than
0.1~meV/atom). Table~\ref{tab:acu} shows that
the relaxed lattice parameters for GaN and InN  in both the wurtzite and 
zinc-blende structures are in good
agreement with other theoretical and experimental work.  
The wurtzite GaN (InN) structures have lower total energies than zinc-blende
GaN (InN) by 13~meV/cation (18~meV/cation).  This is in good agreement
with the calculations of Groose and Neugebauer, where the wurtzite GaN (InN)
is lower in total energy than the zinc-blende GaN (InN) by 11~meV/cation
(19~meV/cation).

\begin{figure}
\resizebox*{3.0in}{!}{\includegraphics[clip]{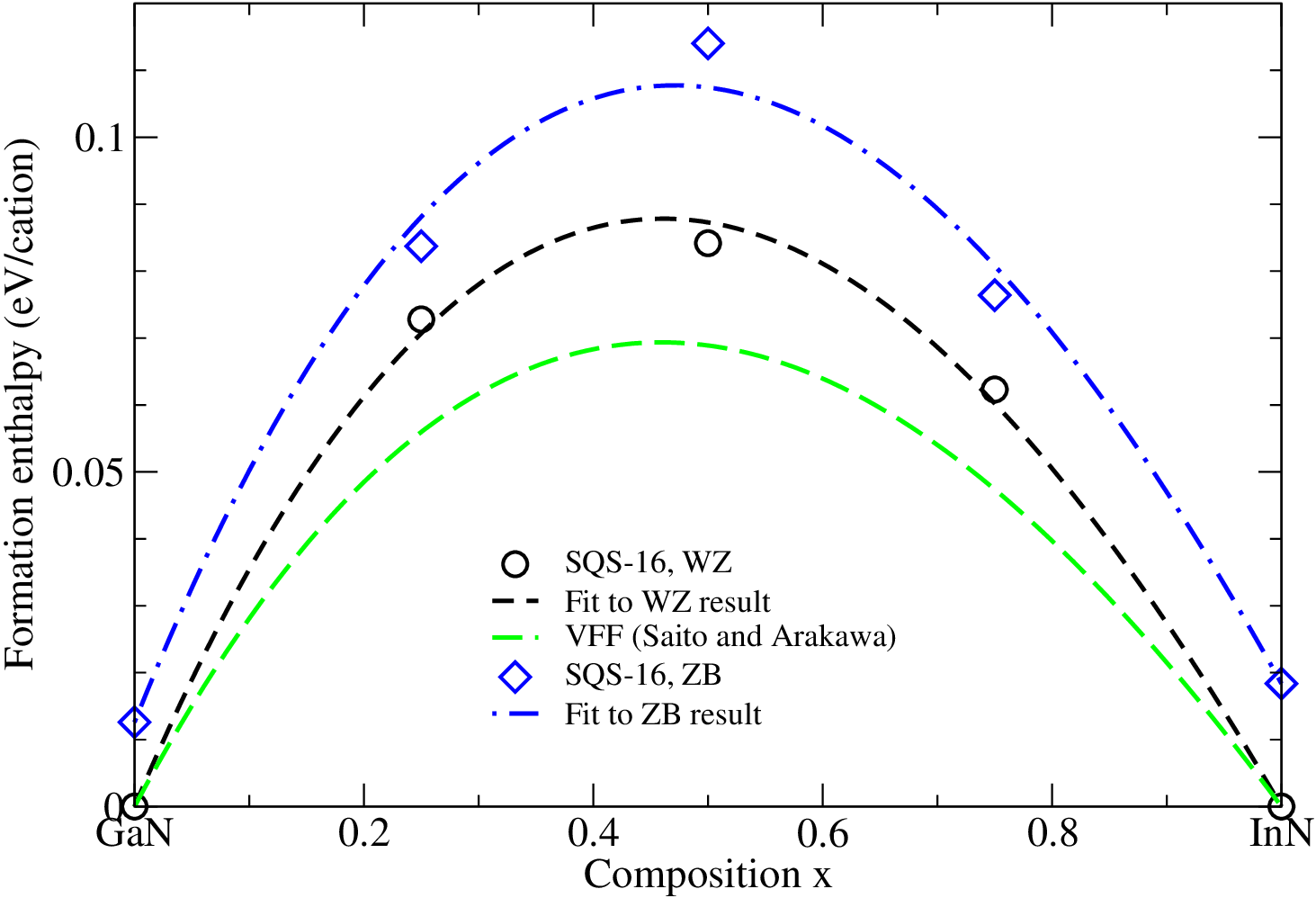}}
\caption{
(Color online)
The formation enthalpy of wurtzite (WZ) and zinc-blende (ZB)
\InGaN\ as a function of 
composition $x$. The solid curve is a fit to the 
calculated values (open circles) with $(\alpha + \beta x)x(1-x)$.
The valence force field results of Saito and Arakawa~\cite{Saito99v60} 
are also included for comparison.
}
\label{fig:wz-zb-formation-enthalpy}
\end{figure}

We determine the equilibrium structure by minimizing the total energy
of the system with respect to the lattice parameters and the positions of 
each of the 32 ions within the SQS cell.
The formation enthalpy of the alloy is 
\begin{equation}
\Delta H_Y = E_Y({\rm In}_x{\rm Ga}_{1-x}{\rm N}) -
x E_{\rm WZ}({\rm InN}) - (1-x)E_{\rm WZ}({\rm GaN}) ,
\label{eq:1}
\end{equation}
where $x = 0$, $ 1/4$, $1/2$, $3/4$, or $1$, and where
$Y$ represents either wurtzite (WZ) or zinc-blende (ZB). 
The wurtzite crystals are used as references for both 
the wurtzite and zinc-blende cases, such that the two  
can be easily compared. The
formation enthalpy results are shown in
Fig.~\ref{fig:wz-zb-formation-enthalpy}.
We fit this data to the form\cite{Saito99v60} $ \Delta H = (\alpha +
\beta x)x(1-x)$ (i.e., beyond regular solution theory). For the wurtzite 
structure, we obtain
$\alpha = 0.4050$~eV/cation and $\beta = -0.1117$~eV/cation.  For 
zinc-blende, we find $\alpha=0.4223$~eV/cation and $\beta =-0.1088$~eV/cation.

Consistent with the results of Saito and Arakawa\cite{Saito99v60} 
(see Fig.~\ref{fig:wz-zb-formation-enthalpy}), we find that all of the 
curves are skewed slightly to the left.  
Focusing on the wurtzite results, we see that the valence force field (VFF) predictions for
the heat of formation\cite{Saito99v60} are
significantly smaller than those obtained here (by as much as
15.4~meV/cation at $x=1/2$).  That the
heat of formation obtained from the the VFF is smaller than the first principles
predictions was first noted by Grosse
and Neugebauer\cite{Grosse01v63} (by ~14.2 meV/cation in their case).

\section{Phonon calculations}
\label{sec:phonon}

The phonon contribution to the free energy was determined within the harmonic
approximation using the  supercell force-constant 
method\cite{Frank95v74,Kresse95v32,Parlinski97v78,Ackland97v9} to calculate 
the phonon density of states. 
In this approach, each atom is displaced
and the forces acting on each of the other (static) atoms are
calculated and used to determine the matrix of force constants.
We then calculate the dynamical matrix
for several reciprocal lattice vectors ${\bf q}$ and diagonalized these to find
the phonon eigen-frequencies $\omega$ and vectors.  Integrating the
phonon eigen-frequencies over the 
Brillouin zone yields the phonon density of states $g(\omega)$. 
At atmospheric pressure, the  difference between
the Gibbs and Helmholtz free energies is negligible for the condensed
phase. 
Therefore, we calculate the phonon free energy $G_v$ for 
an $N$-atom supercell as a function of temperature $T$ as
\begin{equation}
G_v(T) =  N k_B T \int_0^{\infty} d\omega\ g(\omega) 
\log\left[ 2\sinh\left(\frac{\hbar\omega}{2k_B T}\right) \right],
\label{eq:G}
\end{equation}
where $g(\omega)$ is 
normalized according to $\int_0^{\infty} d\omega\ g(\omega) = 3$.

The displacements of the atoms used to calculate the
force constants must be small enough such that the system behaves
harmonically yet large enough that the forces are sufficiently large
as to be reliable.  In our implementation of the vibrational thermodynamics code, 
we employ a displacement of
0.03~\AA\cite{Parlinski99v60}.  Because of the low symmetry inherent
to the random solid solution, each atom in the SQS unit cell is 
displaced individually. For the wurtzite structure, we use a supercell of 64 atoms 
(twice as large as the 16 cation SQS) to reduce the effect of
periodic images for the determination of force constants. For GaN and InN in the 
wurtzite structure, we use a supercell consists of $3\times 3 \times 3 $
four-atom wurtzite primitive cell. For the zinc-blende
structure, we use 64-atom supercells for all
$x$.  The LO/TO splitting, associated with long range dipole-dipole
interactions affects a small region near the center of the Brillouin
zone, was not considered here (for computational efficiency).  This
should introduce only very small errors in the free energy of
formation of the alloy\cite{vandeWalle_error_comment}.  It is possible
to account for the LO/TO splitting using density functional
perturbation theory (DFPT)\cite{Giannozzi91v43,Baroni01v73}.

\begin{figure}
\resizebox*{3.0in}{!}{\includegraphics[clip]{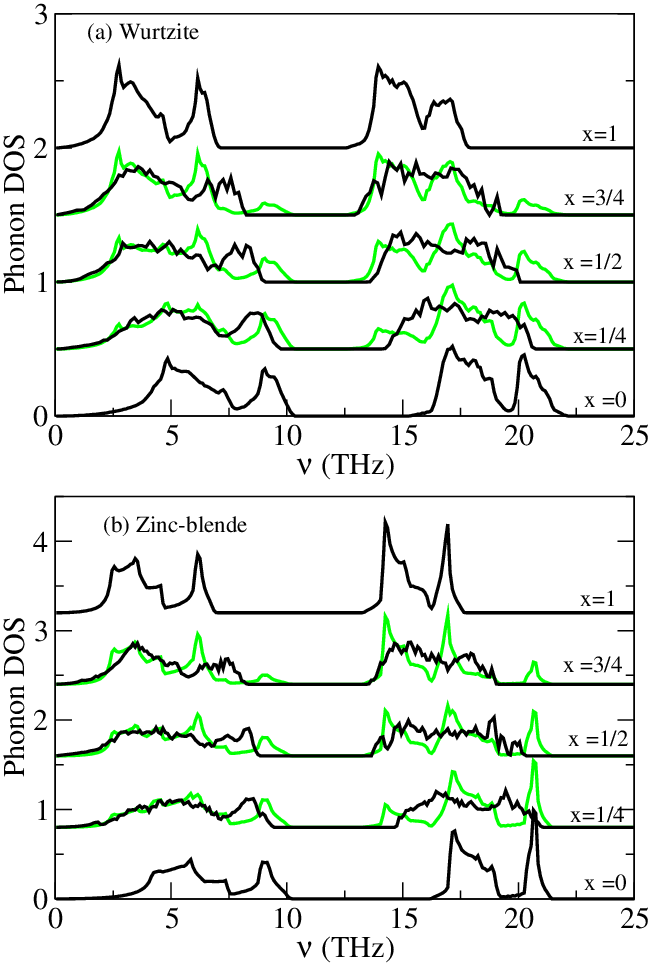}}
\caption{
(Color online) Phonon density of states (DOS) of (a) wurtzite and (b) zinc-blende 
\InGaN\ correspond to $x = 0$, $1/4$,
$1/2$, $3/4$, and $1$, which have been normalized according to
$\int_0^{\infty}g(\nu) d\nu = 3$, where $\nu = \omega/2\pi$.  An
arbitrary shift is added to the curves for clarity. The black and grey curves
refer to the solid solution system $g_{{\rm In}_x {\rm Ga}_{(1-x)}{\rm N}}(\omega)$ 
and an average of the InN and GaN 
phases $g_{\rm av}(\omega)$, respectively.
}
\label{fig:wzzb-phononDOS}
\end{figure}

The calculated phonon densities of states of both wurtzite and 
zinc-blende \InGaN\ are shown in
Fig.~\ref{fig:wzzb-phononDOS} for several alloy compositions. The 
density of states shifts to lower frequencies as the
indium concentration is increased, as expected since (1) In atoms have a
larger mass than Ga atoms and (2) InN has a lower\cite{Wright95v51}
bulk modulus (1.39~Mbar) than GaN (2.02~Mbar).  These two effects 
are of the same magnitude and together explain most of the shift in the vibrational 
frequencies in going from GaN to InN.  Alloying InN with Ga (or vice versa)
decreases the  magnitudes of the peaks in the DOS while broadens the
frequency range.  The broadening of the phonon spectra on alloying
is associated with disorder reducing the coherence of the phonon modes.

\begin{figure}
\resizebox*{3.0in}{!}{\includegraphics[clip]{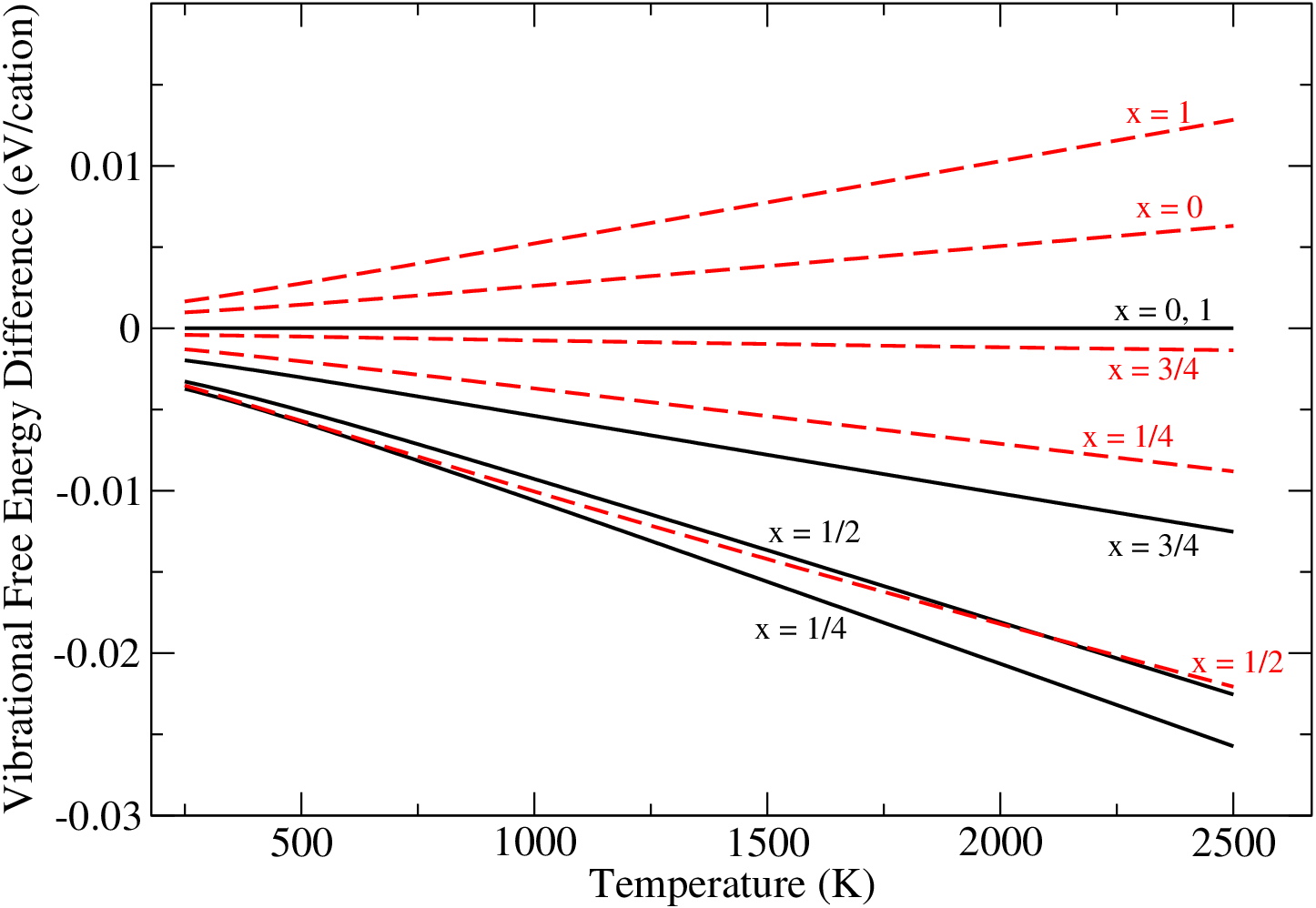}}
\caption{
(Color online)
Vibrational free energy difference $\Delta G_v$ as a function 
of temperature, for the wurtzite (solid lines) and zinc-blende (dashed lines) structures.
}
\label{fig:deltaGv}
\end{figure}

We calculate the vibrational free
energy difference $\Delta G_{v,Y}$ ($Y$ is ZB or WZ) as
\begin{equation}
\Delta G_{v,Y} = G_{v,Y}({\rm In}_x{\rm Ga}_{1-x}{\rm N}) -
x G_{v,{\rm WZ}}({\rm InN}) - (1-x)G_{v,{\rm WZ}}({\rm GaN}),
\label{eq:2}
\end{equation}
where the same wurtzite reference frame is used.
Figure~\ref{fig:deltaGv} shows 
that $\Delta G_{v,Y}$ is a nearly linear function of
temperature for all compositions.
For (nearly) all compositions, $\Delta G_{v,Y}(x)$ for the wurtzite 
structure lies below those for the zinc-blende.  
The vibrational contribution to the entropy of formation 
($\Delta S_v = -\partial \Delta G_v/\partial T$) is larger for the 
alloys ($0<x<1$) than for $x=0$ or $1$ (the negative 
$\Delta S_{v,{\rm ZB}}$  at some compositions is 
an artifact of using wurtzite as the reference structure).  
We note that $\Delta G_v$ for wurtzite In$_{1/4}$Ga$_{3/4}$N is 
lower than that for In$_{1/2}$Ga$_{1/2}$N. 
However, the $\Delta G_v$ for zinc-blende In$_{1/4}$Ga$_{3/4}$N is higher 
than that for zinc-blende In$_{1/2}$Ga$_{1/2}$N. 
We can fit $\Delta G_{v,Y}(x)$ to the same type of interpolation function 
as used for the enthalpy of formation, where the parameters $\alpha$ and $\beta$
in the fit are functions of temperature.

According to Eq.~\ref{eq:2}, $\Delta G_{v}(x)$ is the difference in the 
vibrational contribution to the free energy between the solid 
solution alloy and the terminal compounds.  
This difference can be traced to the difference in phonon density of states 
(Fig.~\ref{fig:wzzb-phononDOS} and Eq.~\ref{eq:G}) between the alloy 
$g_{{\rm In}_x {\rm Ga}_{(1-x)}{\rm N}}(\omega)$ and the composition weighted average 
of the density of states of GaN and InN;  i.e., 
$g_{\rm av}(\omega)=x g_{\rm InN}(\omega)+(1-x) g_{\rm GaN}(\omega)$.
Comparing $g_{{\rm In}_x {\rm Ga}_{(1-x)}{\rm N}}(\omega)$ and $g_{\rm av}(\omega)$
in Fig.~\ref{fig:wzzb-phononDOS},
we see that $g_{\rm av}(\omega)$ is on balance shifted to larger $\omega$ 
at each $x$ (except $x=0$ or $1$, of course).
The density of states at large frequencies $\omega$ makes a larger (more positive)
contribution to the vibrational free energy than do those at small frequencies.
This explains why lattice vibrations lead to negative values of $\Delta G_{v}$. 
We return to this point below.

\section{Pseudo-binary Phase Diagram}
\label{sec:pd}

The free energy of formation $\Delta G$ of the \InGaN\ is given by
\begin{equation}
\Delta G = \Delta H - T\Delta S + \Delta G_{v}.
\end{equation}
As described above, $\Delta H$ and $\Delta G_{v}$ in this equation are 
known from fitting the DFT and phonon 
calculations to particular functional forms.  
$\Delta S$ corresponds to the entropy of mixing and, in the present analysis,
is described in the Bragg-Williams approximation as 
$\Delta S = - k_B [x \log x + (1-x)\log(1-x)]$ (per cation).  
As is well-known, the Bragg-Williams approximation can be severe.
However, the resulting errors are important, predominantly in 
frustrated systems and systems where the phase diagram has a complex
topology\cite{deFontaine79v34}.  Neither of these situations apply to the 
\InGaN\ system.  The Bragg-Williams approximation was also applied
in many of the earlier calculations of the \InGaN\ phase diagram discussed
above\cite{Ho96v69,Saito99v60,Takayama00v88,vanSchilfgaarde97v178}.

\begin{figure}
\resizebox*{3.0in}{!}{\includegraphics[clip]{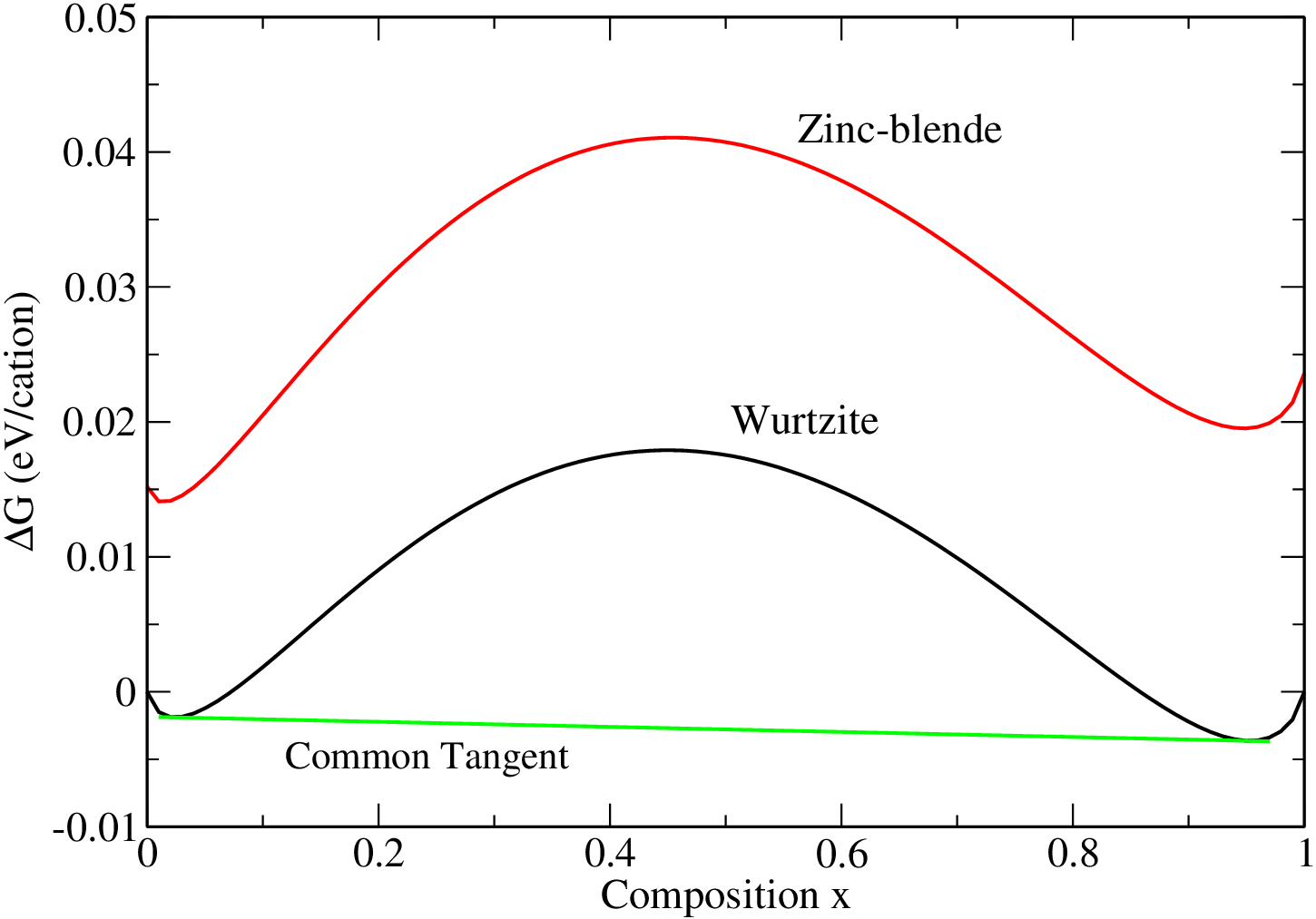}}
\caption{
(Color online)
The common tangent intersects the wurtzite $\Delta G$
curve at $x_1 = 0.024$
and $x_2 = 0.95$, which gives the binodal
curve in Fig.~\ref{fig:wz-w-wo-spin-binodal}. Lattice vibrational 
effect has been taken into account.  A representative case of
$T = 1000$~K is used.
}
\label{fig:wz-1000KCommonTanget}
\end{figure}

We employ the same wurtzite reference state for the unmixed GaN and InN in
order to make a meaningful comparison of the free energies of \InGaN.  Figure
~\ref{fig:wz-1000KCommonTanget} shows $\Delta G$ for both the wurtzite and 
zinc-blende structures. The wurtzite structure has a lower free energy than the 
zinc-blende structure at $T=1000$~K by approximately 0.15 eV/cation.
Further numerical investigation reveals that wurtzite structure has a smaller
free energy than the zinc-blende structure at all 
temperatures and composition $x$
investigated.  This suggests that the wurtzite structure is always more stable than 
the zinc-blende structure in this alloy system, in agreement with experimental
observations.

\label{sec:results}

\begin{figure}
\resizebox*{3.0in}{!}{\includegraphics[clip]{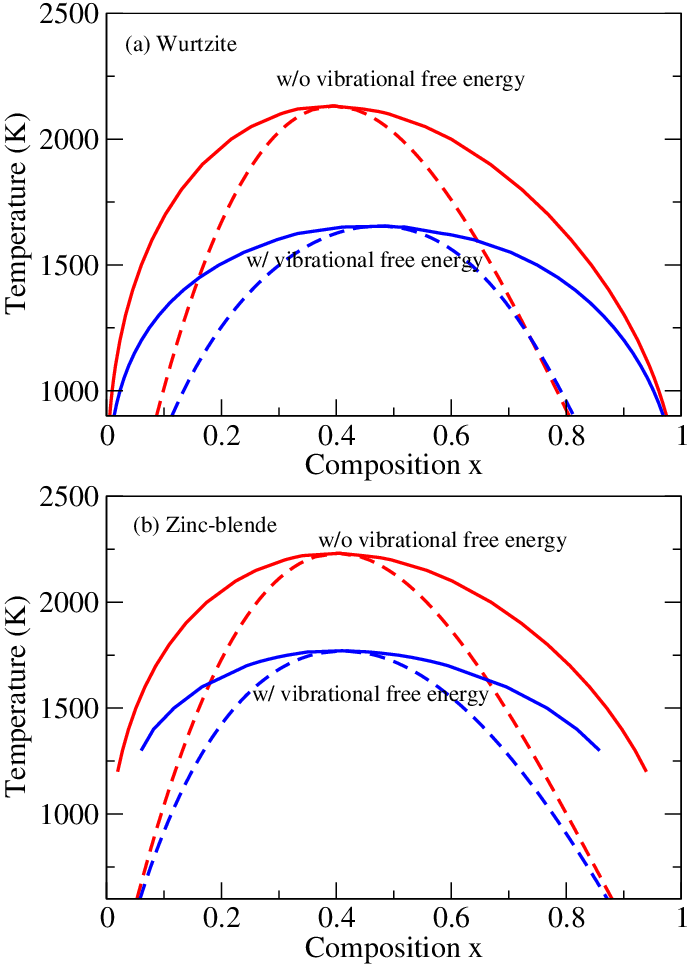}}
\caption{
(Color online) The pseudo-binary diagram (in temperature-cation concentration space) for 
(a) wurtzite and (b) zinc-blende \InGaN\ with and without the inclusion of lattice vibration effects. 
The solid lines correspond to the equilibrium binodal curves and the dashed 
lines represent the spinodal curves. 
}
\label{fig:wz-w-wo-spin-binodal}
\end{figure}

The equilibrium solubility limits (the binodal curve) as a function
of temperature are calculated using a common tangent approach\cite{Kittel80}, as
shown in Fig.~\ref{fig:wz-1000KCommonTanget} for 1000 K. 
Figure~\ref{fig:wz-w-wo-spin-binodal} shows that the critical temperature
below which phase separation will occur at some composition,   
$T_c$, is 2132~K for the wurtzite structure in the absence of 
vibrational contributions to the free energy.
Zinc-blende \InGaN\ shows $T_c = 2231$~K when 
vibrational effects are excluded.
These values of $T_c$ (for both the wurtzite and zinc-blende structures) are 
higher than other VFF results\cite{Ho96v69,Saito99v60,Takayama00v88}
that neglect vibrations.  These differences are 
attributable to the fact that the 
formation enthalpy determined from our DFT calculations are  
larger than those obtained within the VFF 
(e.g., see Fig.~\ref{fig:wz-zb-formation-enthalpy}).
Our predicted $T_c = 2231$~K (for zinc-blende structure without vibrations) 
is considerably higher than those obtained by Teles {\it et al.}\cite{Teles00v62} 
($T_c  = 1295$~K).  This is probably attributable to the 
relatively small (8-atom) supercells that they employed. 
We note that another DFT calculation\cite{Ferhat02v65}, 
with larger clusters (64 atoms/cell), predicts a slightly higher 
$T_c = 1400$~K. The discrepancy that still remains is likely the result 
of the better description of the random solid solution alloy used in our
analysis. 

Inclusion of the lattice vibrational effect
reduces (see Fig.~\ref{fig:wz-w-wo-spin-binodal}) 
$T_c$ for the wurtzite \InGaN\ from 2132~K to 1654~K, a decrease
of 29\%. For zinc-blende structure, $T_c$ is reduced from 2231~K to
1771~K. 
The change in the critical temperature upon
inclusion of the vibrational contributions suggests that omission of
this term may be one of the most significant errors commonly made
in the calculation of phase diagrams of compound semiconductor
alloys. Since this contribution is largely entropic and is larger in the
disordered phase, it will generally lead to a decrease in the
stability of the ordered phase relative to the disordered phase and, therefore, to a
lower $T_c$. The origin of this effect can be traced to the broadening
of the vibrational density of state with increasing disorder, as shown
in Fig.~\ref{fig:wzzb-phononDOS}.
Inclusion of the vibrational
contribution to the free energy also makes the 
wurtzite \InGaN\ phase diagram more
symmetric with a peak at $x=0.47$ with vibrations, as compared to 
$x=0.39$ without. The maximum in the binodal in the zinc-blende diagram also 
shifts closer to $x=0.5$ with inclusion of vibrational effects (although in this
case the effect is weaker than in the wurtzite case). 
We also note that inclusion of vibrations does not 
significantly increase the solubilities of In in GaN or Ga in InN, 
except above approximately 1200~K.

Another effect of vibrations is to decrease the width of the 2-phase field
in the phase diagram at all temperatures (relative to the phase diagram
that does not include the vibration effect).  A common goal in the growth
of  \InGaN\ alloys for microelectronic applications is to increase the
range of compositions that are achievable without phase separation (i.e.,
increase the solubility limits).  Examination of the phase diagram  in
Fig.~\ref{fig:wz-w-wo-spin-binodal} shows that this could be
accomplished simply by increasing the growth temperature.  This is
not always possible.  An alternative approach is to increase the entropy
of the disordered alloy relative to that of the phase separated material.
This could be accomplished through alloying (quaternary additions).
Addition of a new
component increases the configurational entropy of the alloy.  The
present results suggests that alloying could also be used to increase the
vibrational contribution to the stability of the disordered phase.  This
would be most effective with the addition of heavier elements, which shift
the vibrational frequencies to lower values and therefore increase the
width of the vibrational density of state (see Fig.~\ref{fig:wzzb-phononDOS}).  Of
course, the new alloy component should not increase the enthalpy of
formation of the disordered phase.

\begin{figure}
\resizebox*{3.0in}{!}{\includegraphics[clip]{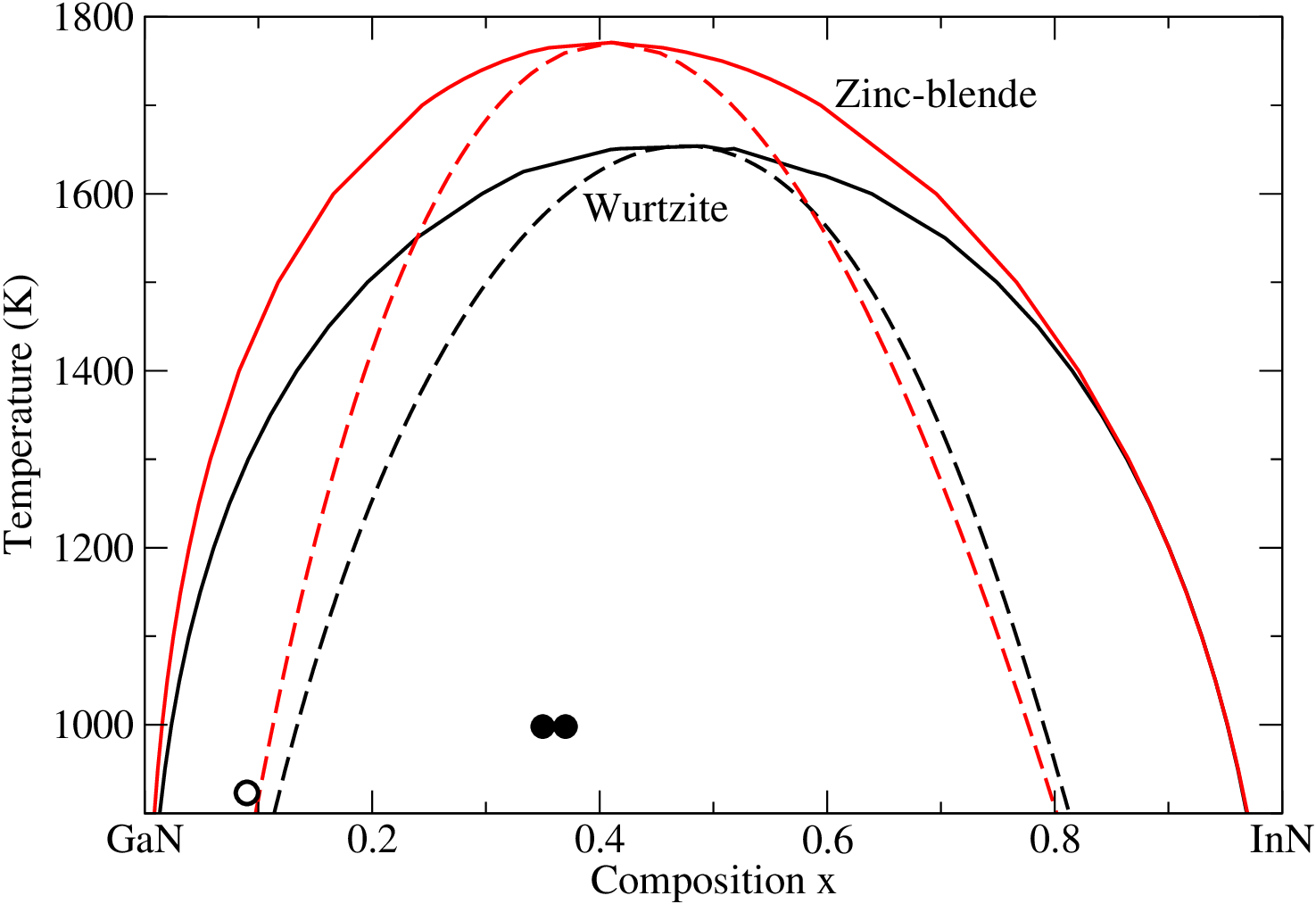}}
\caption{
(Color online) A comparison of the pseudo-binary diagrams of zinc-blende and 
wurtzite \InGaN\ (including the effects of lattice vibrations). 
The solid lines correspond to the equilibrium binodal curves and the dashed 
lines represent the spinodal curves. The points represented by the filled and open circles
represent the experimental data\cite{Doppalapudi98v84} for which phase separation was and was not observed, respectively.
}
\label{fig:wz-zb-bi-spin}
\end{figure}

Figure~\ref{fig:wz-zb-bi-spin} shows a comparison of the wurtzite and
zinc-blende \InGaN\ phase diagrams (including the effects of lattice 
vibrations). The wurtzite structure has a lower $T_c$ and is more 
symmetric (with respect to composition) than is the zinc-blende structure.
These differences can be important in cases where the zinc-blende 
structure is stabilized through heteroepitaxy\cite{Taniyasu00v180,Tabata99v75}
or other means.

A direct comparison of the predicted phase diagram with an experimental 
phase diagram cannot be made since the latter has never been determined.
However, an experimental study\cite{Doppalapudi98v84} of
In$_{0.09}$Ga$_{0.91}$N showed that this alloy does not phase 
separate at the growth temperature of 923~K. According to the phase 
diagram in Fig.~\ref{fig:wz-zb-bi-spin}, this composition and temperature
lies between the binodal and spinodal.  This suggests that the solid solution
is metastable, with respect to phase separation. 
Additional experimental measurements\cite{Doppalapudi98v84} show that phase separation occurs
in samples grown at both In$_{0.37}$Ga$_{0.63}$N and 
In$_{0.35}$Ga$_{0.65}$N and 998~K.  This is 
also consistent with the predicted phase diagram --- these growth conditions
are within the spinodal region, where the disordered alloy is unstable.  
Since these materials were produced by MOCVD, there is some uncertainty
as to whether the phases observed correspond to the bulk equilibrium 
phases at this temperature.  In such cases, the phases that are observed
may be influenced/inherited from the surface structure and by hetero-epitaxial
strains.  However, a solid solution cannot be stable within the spinodal region 
of the phase diagram, independent of how the material was produced 
(slow diffusional kinetics may limit the rate at which such an instability grows).

The phase diagrams presented here constructed based upon several
assumptions.  These include a harmonic description of the phonons in a 
relatively small unit cell, a Bragg-Williams approximation for the entropy of 
mixing and that the solid solution has no short-range order. 
All approximations employed in this work can be 
improved by using larger unit cells, a quasi-harmonic model for the vibrational
contributions to the free energy, more accurate treatment of the entropy of mixing
(e.g., using a cluster variational approach\cite{Zunger94}) and a self-consistent
determination of the short-range order.  Despite the nature of the approximations
made, the predicted results represent the most rigorous determination of the
\InGaN\ phase diagram to date.

\section{Summary}
\label{sec:summary}
We have employed accurate density-functional calculations
to study the thermodynamic properties of both the zinc-blende and wurtzite
structures of \InGaN. The disordered alloy was modeled using the quasi-random 
structure (SQS) approach to ensure that the order parameter accurately 
describes a random solid solution (the detailed SQS structure information is 
provided for the 16-cation unit cells for both the zinc-blende and wurtzite 
structures). The vibrational contribution to the free energy has been calculated 
using first-principles calculations of the phonon spectra within the harmonic 
approximation. Inclusion of these effects leads to a 29\% (26\%) reduction 
in the critical temperature for phase separation in the wurtzite (zinc-blende) 
structure. We found that the wurtzite structure is thermodynamically more stable
than zinc-blende structure at all temperatures and alloy compositions.  The 
predicted miscibility gap and spinodal curve are consistent with experimental
observations on MOCVD \InGaN.  Nonetheless, care must be exercised in
interpreting the phases observed following such growth processes in terms
of an equilibrium phase diagram.

The present results suggest that inclusion of the vibrational
contribution to the free energy of formation of compound semiconductor
alloys can have a profound effect on the critical temperature for the
order-disorder transition on the cation sublattice.  This
issue is particularly important for the increasingly complex compound
semiconductors of interest today.  Not only does the
vibrational entropy shift the critical temperature, it also modifies
the solubility limits of each component and does so in a non-symmetric
manner.  The present results also point to the importance of
determining all of the contributions to the free energy
self-consistently; that is using the same energetics description for
relaxing the atomic structures, calculating enthalpies of formation
and vibrational contributions to the free energy. While the present
results represents a significant advance in compound semiconductor
alloy phase diagram calculations, further improvements should include
the effects of short range order in the disordered phase and better
approximations to the entropy of mixing (see e.g.,
Ref.~\onlinecite{Zunger94}).

\section{acknowledgments}
The authors gratefully acknowledge useful discussions with
C.~Jiang, A.~van de Walle and D. Vanderbilt.
This work was supported by Visiting Investigator Program, Agency for 
Science, Technology and Research (A*STAR), Singapore. 

\clearpage\pagebreak

\begin{table}[p]
\caption{The SQS wurtzite (hcp) structure with 16 cations and 16 anions 
for A$_x$B$_{1-x}N$ alloys where A and B represent In and Ga cations, 
respectively, with $x= 1/4$ (or $x=3/4$)
and $x=1/2$.  The lattice vectors ${\bf a}_1$, ${\bf a}_2$, and ${\bf
a}_3$ of the structures are linear combinations of the lattice vectors
(i.e., ${\bf a}$, ${\bf b}$, and ${\bf c}$) of a hexagonal primitive
cell.  In Cartesian coordinates ${\bf a} = (a_0, 0, 0)$, ${\bf b} =
(-a_0/2, a_0\sqrt{3}/2,0)$ and ${\bf c} = (0, 0, c_0)$. The positions
of the A cations are  represented in the table as A($\alpha$,$\beta$,
$\gamma$) corresponding to an A cation at the position
($\alpha{\bf a}_1+\beta {\bf a}_2+\gamma {\bf a}_3)/48$.}

\begin{center}
\begin{tabular}{l|l}
\hline\hline
$x = 1/4$ & $x = 1/2$ \\
\hline
 ${\bf a}_1 = 2{\bf a} -2 {\bf b} -2 {\bf c}$ & ${\bf a}_1 = 2{\bf a} +2 {\bf b} -2 {\bf c}$ \\
${\bf a}_2 = {\bf a} +2 {\bf b}$  & ${\bf a}_2 = 2{\bf a} +  {\bf b}- {\bf c}$ \\
${\bf a}_3 = {\bf a} + {\bf b} + {\bf c}$ &  ${\bf a}_3 = 2{\bf a} - 2 {\bf b} - 2 {\bf c}$\\
A$(39,44,6)$, A$(18,40,36)$ & A$(12,16,28)$, A$(30, 8,26)$ \\
A$(24,16,0)$, A$(27,44,30)$  & A$(24,16,16)$, A$(42, 8,14)$ \\
B$( 0,16, 0)$, B$( 3,44,30)$& A$(42, 8,38)$, A$(30, 8, 2)$\\
B$( 6,40,12)$, B$( 9,20,42)$& A$(36,16,28)$, A$(  0,16,16)$ \\
B$(42,40,36)$, B$(45,20,18)$& B$(  0,16,40)$, B$(18, 8,38)$\\
B$(12,16,24)$, B$(15,44, 6)$& B$(36,16, 4)$, B$( 6, 8, 2)$\\
B$(36,16,24)$, B$(21,20,18)$ & B$(24,16,40)$, B$(12,16, 4)$\\
B$(30,40,12)$, B$(33,20,42)$ & B$( 6, 8,26)$, B$(18, 8,14)$ \\
\hline\hline
\end{tabular}
\end{center}
\label{tab:wurtzitesqs16}
\end{table}

\begin{table}[p]
\caption{The 16 cation SQS zinc-blende (fcc) structure 
for A$_x$B$_{1-x}N$ alloys.  The lattice vectors ${\bf a}_1$, ${\bf a}_2$, and ${\bf
a}_3$ are reported in terms of the lattice vectors of the primitive rhombohedral 
cell:  ${\bf a} = a_0(0, 1/2,
1/2)$, ${\bf b} = a_0(1/2, 0, 1/2)$ and ${\bf c} = a_0(1/2, 1/2,
0)$.  The positions of the A cations are represented in the table as A($\alpha$,$\beta$,
$\gamma$) corresponding to an A cation at the position
($\alpha{\bf a}_1+\beta {\bf a}_2+\gamma {\bf a}_3)/16$.}

\begin{center}
\begin{tabular}{l|l}
\hline\hline
$x = 1/4$ & $x = 1/2$ \\
\hline
 ${\bf a}_1 = 2{\bf a} + {\bf b} -2 {\bf c}$ & ${\bf a}_1 = 2{\bf a} -2 {\bf c}$ \\
${\bf a}_2 = {\bf a} + {\bf b} + 2 {\bf c}$   & ${\bf a}_2 = 2{\bf a} +  2 {\bf b}$ \\
${\bf a}_3 = 2 {\bf a} - {\bf b} + 2 {\bf c}$ &  ${\bf a}_3 = 2{\bf a} - 2 {\bf b}$\\
A$(11,12,7)$, A$(4,0,4)$ &     A$(8,4,4)$, A$(0,8,8)$  \\
A$(15,12,11)$, A$(9,4,13)$ &  A$(8,0,8)$, A$(0,4,12)$  \\
B$(0,0,0)$, B$(3,12,15)$ &       A$(8,0,0)$, A$(0,4,4)$  \\
B$(13,4,1)$, B$(6,8,14)$ &    A$(8,12,4)$, A$(0,0,8)$  \\
B$(10,8,2)$, B$(14,8,6)$ &    B$(0,0,0)$, B$(8,12,12)$  \\
B$(1,4,5)$, B$(8,0,8)$ &       B$(0,12,4)$, B$(8,8,0)$  \\
B$(2,8,10)$, B$(5,4,9)$ &     B$(0,12,12)$, B$(8,8,8)$  \\
B$(12,0,12)$, B$(7,12,3)$ &    B$(0,8,0)$, B$(8,4,12)$  \\
\hline\hline
\end{tabular}
\end{center}
\label{tab:zinc-blendesqs16}
\end{table}

\begin{table}[p]
\caption{Lattice parameters of the wurtzite (WZ) and 
zinc-blende (ZB) structures for 
GaN and InN from the 
present study and other studies.  $u$
describes the displacement of the N (0001) planes from the Ga (0001)
planes in units of $c$ as described, in detail, in e.g.,
Ref. \onlinecite{Burns85}.}
\begin{center}
\begin{tabular}{ccccc}
\hline\hline
& & $a$(\AA) & $c$(\AA) & $u$ \\
\hline
GaN (WZ) & Present calc. & 3.145 & 5.121 & 0.377 \\
& Calc.\cite{Grosse01v63} & 3.196 & 5.206 & 0.376 \\
& Calc.\cite{Bungaro00v61} & 3.20 & 5.22 & 0.376 \\
& Calc.\cite{Parlinski99v60} & 3.133 & 5.108 & 0.377 \\
& Calc.\cite{Wright94v50} & 3.162 & 5.142 & 0.377\\
& Expt.\cite{Schulz77v23} & 3.190 & 5.189 & 0.377 \\
\hline
InN (WZ) & Present calc. & 3.518 & 5.690 & 0.379 \\
& Calc.\cite{Grosse01v63} & 3.545 & 5.761 & 0.376 \\
& Calc.\cite{Bungaro00v61} & 3.48 & 5.64 & 0.378 \\
& Calc.\cite{Wright95v51} & 3.501 & 5.669 & 0.3784\\
& Expt.\cite{Osamura75v46} & 3.544 & 5.718 &  \\
\hline
GaN (ZB) & Present calc. & 4.443 &  &  \\
& Calc.\cite{Karch98v57} & 4.447 &  &  \\
\hline
InN (ZB) & Present calc. & 4.964 & & \\
& Calc.\cite{Wright95v51} & 4.932 &  &  \\
\hline\hline
\end{tabular}
\end{center}
\label{tab:acu}
\end{table}

\clearpage\pagebreak



\end{document}